\documentclass[12pt]{article}
\usepackage{graphicx}
\usepackage{color}

\def\hybrid{\topmargin 0pt      \oddsidemargin 0pt
        \headheight 0pt \headsep 0pt
       \voffset-1cm
        \textwidth 6.25in       
       \textheight 9.5in       
        \marginparwidth 0.0in
        \parskip 5pt plus 1pt   \jot = 1.5ex}
\catcode`\@=11
\def\marginnote#1{}

\newcount\hour
\newcount\minute
\newtoks\amorpm
\hour=\time\divide\hour by60
\minute=\time{\multiply\hour by60 \global\advance\minute by-\hour}
\edef\standardtime{{\ifnum\hour<12 \global\amorpm={am}%
        \else\global\amorpm={pm}\advance\hour by-12 \fi
        \ifnum\hour=0 \hour=12 \fi
        \number\hour:\ifnum\minute<10 0\fi\number\minute\the\amorpm}}
\edef\militarytime{\number\hour:\ifnum\minute<10 0\fi\number\minute}

\def\draftlabel#1{{\@bsphack\if@filesw {\let\thepage\relax
   \xdef\@gtempa{\write\@auxout{\string
      \newlabel{#1}{{\@currentlabel}{\thepage}}}}}\@gtempa
   \if@nobreak \ifvmode\nobreak\fi\fi\fi\@esphack}
        \gdef\@eqnlabel{#1}}
\def\@eqnlabel{}
\def\@vacuum{}
\def\draftmarginnote#1{\marginpar{\raggedright\scriptsize\tt#1}}

\def\draftlabel#1{{\@bsphack\if@filesw {\let\thepage\relax
   \xdef\@gtempa{\write\@auxout{\string
      \newlabel{#1}{{\@currentlabel}{\thepage}}}}}\@gtempa
   \if@nobreak \ifvmode\nobreak\fi\fi\fi\@esphack}
        \gdef\@eqnlabel{#1}}
\def\@eqnlabel{}
\def\@vacuum{}
\def\draftmarginnote#1{\marginpar{\raggedright\scriptsize\tt#1}}

\def\draft{\oddsidemargin -.5truein
        \def\@oddfoot{\sl preliminary draft \hfil
        \rm\thepage\hfil\sl\today\quad\militarytime}
        \let\@evenfoot\@oddfoot \overfullrule 3pt
        \let\label=\draftlabel
        \let\marginnote=\draftmarginnote
   \def\@eqnnum{(\theequation)\rlap{\kern\marginparsep\tt\@eqnlabel}%
\global\let\@eqnlabel\@vacuum}  }

\def\numberbysection{\@addtoreset{equation}{section}
        \def\theequation{\thesection.\arabic{equation}}}

\def\underline#1{\relax\ifmmode\@@underline#1\else
        $\@@underline{\hbox{#1}}$\relax\fi}

\def\titlepage{\@restonecolfalse\if@twocolumn\@restonecoltrue\onecolumn
     \else \newpage \fi \thispagestyle{empty}\c@page\z@
        \def\thefootnote{\fnsymbol{footnote}} }

\def\endtitlepage{\if@restonecol\twocolumn \else  \fi
        \def\thefootnote{\arabic{footnote}}
        \setcounter{footnote}{0}}  
\relax

\numberbysection
\hybrid

\newfont{\Bbb}{msbm10 scaled 1\@ptsize00}
\newfont{\Bbbb}{msbm7 scaled 1\@ptsize00}

\newcommand{\DDD}{\raise-1pt\hbox{$\mbox{\Bbbb D}$}}

\newcommand{\UUU}{\raise-1pt\hbox{$\mbox{\Bbbb U}$}}

\newcommand{\z}{\raise-1pt\hbox{$\mbox{\Bbbb Z}$}}

\def\res{\mathop{\hbox{res}}\limits}

\def\beq{\begin{equation}}
\def\eeq{\end{equation}}
\def\p{\partial}

\begin{document}

\begin{titlepage}

\title{KP hierarchy and trigonometric Calogero-Moser hierarchy}

\author{A.~Zabrodin\thanks{Skolkovo Institute of Science and Technology, 143026 Moscow, Russian Federation;
Institute of Biochemical Physics, Kosygina str. 4, 119334, Moscow, Russian Federation;
e-mail: zabrodin@itep.ru}
}

\date{June 2019}
\maketitle

\vspace{-7cm} \centerline{ \hfill ITEP-TH-14/19}\vspace{7cm}

\begin{abstract}

We consider trigonometric solutions of the KP hierarchy. It is known that 
their poles
move as particles of the Calogero-Moser model with trigonometric potential. 
We show that this correspondence can be extended to the level of hierarchies: the evolution
of the poles with respect to the $k$-th hierarchical time of the KP hierarchy is governed by a 
Hamiltonian which is a linear combination of the first $k$ higher Hamiltonians 
of the trigonometric Calogero-Moser hierarchy.

\end{abstract}

\end{titlepage}


\vspace{5mm}

\section{Introduction}

The Kadomtsev-Petviashvili (KP) hierarchy 
is an archetypal object in the theory of integrable systems. It is an infinite set
of compatible nonlinear partial differential equations involving infinitely many 
variables ${\bf t}=\{t_1, t_2, t_3, \ldots \}$ (``hierarchical times''). These equations
admit a huge number of solutions of very different nature. Among them, of special
interest are singular solutions with a finite number of moving poles. In particular,
one can consider solutions for which the dependent variables are trigonometric (or hyperbolic)
functions of $t_1=x$ with poles depending on the times $t_k$ with $k\geq 2$ (``trigonometric solitons''). 

Dynamics of poles of singular solutions to nonlinear integrable equations is a rather
familiar subject in the theory of integrable systems. These studies were initiated by 
the seminal paper \cite{AMM77}, where elliptic and rational solutions to the Korteweg-de
Vries and Boussinesq equations were investigated. A remarkable connection with integrable
many-body Calogero-Moser systems \cite{Calogero71,Calogero75,Moser75,OP81} was observed. Later
in \cite{Krichever78,CC77} it has been shown that this connection becomes most natural for the 
more general KP equation: the evolution of poles of rational solutions to the KP equation
with respect to the time $t_2$ is the Calogero-Moser dynamics of the many-body 
system with the rational pairwise interaction potential
$1/(x_i-x_j)^2$. In \cite{Krichever80} Krichever has extended this connection 
to elliptic (double
periodic) solutions expressed through the Weierstrass $\wp$-function. The method
suggested by Krichever consists in substituting the singular solution not in the KP
equation itself but in the auxiliary linear problem for it, using a suitable pole ansatz
for the wave function. This method allows one to obtain the equations of motion
together with the Lax representation for them. 

The further development is Shiota's work \cite{Shiota94}. Shiota has shown that 
the correspondence between rational solutions to the KP equation and the Calogero-Moser
system with rational potential can be extended to the level of hierarchies: namely,
the evolution of poles with respect to the higher times $t_k$ is governed by 
the higher Hamiltonians $H_k$ of the integrable Calogero-Moser system. The higher 
Hamiltonians are defined as traces of powers of the Lax matrix $L$: 
$H_k=\mbox{tr}\, L^k$. 

Trigonometric and hyperbolic (single-periodic
in the complex plane) solutions to the KP hierarchy are not so well studied up to now. 
In this paper we extend Shiota's method to this class of solutions.
They have the form
$$
u(x, {\bf t})=-\sum_{i=1}^N \frac{\gamma^2}{\sinh ^2 (\gamma (x-x_i({\bf t}))},
$$
where $\gamma$ is a complex parameter. When $\gamma$ is purely imaginary (respectively, real),
one deals with trigonometric (respectively, hyperbolic) solutions. The limit $\gamma \to 0$
corresponds to rational solutions. 
We show that 
the evolution of the poles $x_i$ with respect to the higher times $t_k$ is governed by 
the Hamiltonians
\beq\label{i1}
{\cal H}_k =\frac{1}{2(k+1)\gamma}\, \mbox{tr}
\Bigl ((L\! +\! \gamma I)^{k+1}-(L\! -\! \gamma I)^{k+1}\Bigr )
\eeq
which are linear combinations of the Hamiltonians $H_k =\mbox{tr}\, L^k$.
Here $I$ is the identity matrix and $L$ is the Lax matrix given by
\beq\label{i2}
L_{ij}=-p_i \delta_{ij}-\frac{(1-\delta_{ij})\gamma}{\sinh (\gamma (x_i-x_j))}.
\eeq
In particular, 
\beq\label{i3}
{\cal H}_2=H_2+\mbox{const} = \sum_{i=1}^{N}p_i^2 -\sum_{i\neq j}
\frac{\gamma^2}{\sinh ^2 (\gamma (x_i-x_j))} +\mbox{const}
\eeq
is the Hamiltonian of the trigonometric Calogero-Moser system.
The quantities $H_k$ (and ${\cal H}_k$) are integrals of motion because the evolution
is an isospectral transformation of the Lax matrix.

Our method consists in a direct solution of the auxiliary linear problems
for the wave function and its adjoint using a pole ansatz. A similar result was obtained
in \cite{Haine07} within a different approach. 

The organization of the paper is as follows. In section 2 we very briefly review the KP
hierarchy introducing the main notions of the Lax operator, auxiliary linear problems
and tau-function. Section 3 is devoted to the trigonometric solutions. We derive the dynamics
of their poles in the time $t_2$. In section 4 the dynamics with respect to the higher times
is considered and the Hamiltonian equations for the higher flows are derived. 
In section 5 we consider the B\"acklund transformation of the pole dynamics depending on a 
parameter.
In section 6
we prove the determinant formula for the tau-function for trigonometric 
solutions which expresses it
through the initial values of the coordinates and momenta. In the appendix we give some 
details on the expansion of the B\"acklund transformation in powers of the parameter
which yields equations of motion for the higher times. 

\section{The KP hierarchy}

The KP hierarchy can be understood as an infinite set of evolution equations in the times
${\bf t}$ for functions of a variable $x$. Let
\beq\label{kp1}
{\cal L}=\p_x +\sum_{k\geq 1}u_k \p_x^{-k}
\eeq
be the pseudo-differential Lax operator, where the coefficient functions $u_k$ are
functions of $x$ and ${\bf t}$. The equations of the KP hierarchy for $u_k$'s 
are encoded in the 
Lax equations
\beq\label{kp2}
\p_{t_m}{\cal L}=[{\cal A}_m, {\cal L}], \qquad {\cal A}_m=({\cal L}^m)_+,
\eeq
where $(\ldots )_+$ means taking the purely differential part of a pseudo-differential
operator. In particular, we have $\p_{t_1}{\cal L}=\p_x {\cal L}$, i.e., 
$\p_{t_1}u_k =\p_x u_k$ for all $k\geq 1$. This means that the evolution in $t_1$ is simply
a shift of $x$: $u_k (x, {\bf t})=u_k(x+t_1, t_2, t_3, \ldots )$. 

An equivalent formulation of the KP hierarchy is through the zero curvature
(Za\-kha\-rov-\-Sha\-bat) equations
\beq\label{kp3}
\p_{t_n}{\cal A}_m -\p_{t_m}{\cal A}_n +[{\cal A}_m, {\cal A}_n]=0.
\eeq
The simplest nontrivial equation is obtained at $m=2$, $n=3$. It is the famous
KP equation for $u=u_1$:
\beq\label{kp3a}
3u_{t_2t_2}=\Bigl (4u_{t_3}-12uu_x -u_{xxx}\Bigr )_x.
\eeq
The Zakharov-Shhabat equations
are compatibility conditions for the auxiliary linear problems
\beq\label{kp4}
\p_{t_m}\psi ={\cal A}_m \psi ,
\eeq
where the wave function $\psi$ depends on a spectral parameter $z$: 
$\psi =\psi (z; {\bf t})$. Together with the wave function $\psi$ one can introduce
the adjoint wave function $\psi^{\dag}$ which satisfies the conjugate linear equations
\beq\label{kp5}
-\p_{t_m}\psi^{\dag} ={\cal A}^{\dag}_m \psi^{\dag}
\eeq
The conjugation of a differential operator is performed according to the rule
$(f(x)\p_x^n)^{\dag}=(-\p_x)^n f(x)$.
In particular, we have the auxiliary linear problems
\beq\label{kp6}
\p_{t_2}\psi = \p_x^2\psi +2u_1 \psi , \qquad
-\p_{t_2}\psi^{\dag} = \p_x^2\psi^{\dag} +2u_1 \psi^{\dag}
\eeq
which have the form of the non-stationary Schrodinger equation.

A common solution to the KP hierarchy is provided by the tau-function
$\tau = \tau (x, {\bf t})$. 
The whole hierarchy is encoded in the bilinear relation \cite{DJKM83,JM83}
\beq\label{kp7}
\oint_{\infty}e^{(x-x')z +\xi ({\bf t}, z)-\xi ({\bf t}', z)}
\tau \Bigl (x, {\bf t}-[z^{-1}]\Bigr )\tau \Bigl (x', {\bf t}'+[z^{-1}]\Bigr )dz =0
\eeq
valid for all $x, x'$, ${\bf t}, {\bf t}'$, where
$$
\xi ({\bf t}, z)=\sum_{k\geq 1}t_k z^k,
$$
\beq\label{kp8}
{\bf t}\pm [z^{-1}]=\Bigl \{t_1\pm z^{-1}, t_2\pm \frac{1}{2}z^{-2}, 
t_3\pm \frac{1}{3}z^{-3}, \ldots \Bigl \}.
\eeq
The integration contour is a big circle around infinity separating the singularities
coming from the exponential factor from those coming from the tau-functions. 
A consequence of the bilinear
relation (which is in fact equivalent to the whole hierarchy, see \cite{TT95}) is the equation
\beq\label{kp9}
\begin{array}{c}
\p_x \tau \Bigl (x, {\bf t}+[\lambda^{-1}]\Bigr )
\tau \Bigl (x, {\bf t}+[\mu^{-1}]\Bigr )
- \p_x \tau \Bigl (x, {\bf t}+[\mu^{-1}]\Bigr )
\tau \Bigl (x, {\bf t}+[\lambda^{-1}]\Bigr )
\\ \\
=\, (\lambda -\mu )\Bigl [\tau \Bigl (x, {\bf t}+[\lambda^{-1}]\Bigr )
\tau \Bigl (x, {\bf t}+[\mu^{-1}]\Bigr )-
\tau \Bigl (x, {\bf t}+[\lambda ^{-1}]+ [\mu^{-1}]\Bigr )\tau (x, {\bf t})\Bigr ].
\end{array}
\eeq
In this form this equation appeared for example in \cite{AvanM92,Takasaki07,NZ16}.
The differential equations of the hierarchy are obtained by expanding this equation
in inverse powers of $\lambda$, $\mu$. It is important to note that the tau-functions
which differ by an exponential factor of a linear combination of times are equivalent.

The coefficient functions $u_k$ can be expressed through the tau-function. 
In particular,
\beq\label{kp10}
u_1(x, {\bf t})=\p_x^2\log \tau (x, {\bf t}).
\eeq
The wave function and its adjoint are expressed through the tau-function 
according to the formulas
\beq\label{kp11}
\psi (z; {\bf t})=A(z)\, e^{xz+ \xi ({\bf t}, z)}
\frac{\tau (x, {\bf t}-[z^{-1}])}{\tau (x, {\bf t})},
\eeq
\beq\label{kp12}
\psi^{\dag} (z; {\bf t})=A(z)\, e^{-xz-\xi ({\bf t}, z)}
\frac{\tau (x, {\bf t}+[z^{-1}])}{\tau (x, {\bf t})},
\eeq
where $A(z)$ is a normalization factor. 

Let us point out another useful corollary of the bilinear relation.
Differentiating (\ref{kp7}) with respect to $t_m$ and putting $x=x'$, 
${\bf t}={\bf t}'$
after that, we obtain
\beq\label{kp7a}
\frac{1}{2\pi i}\oint_{\infty}z^m \psi^{\dag}(z; {\bf t})\psi (z; {\bf t})
dz = \p_{t_m}\p_x \log \tau (x, {\bf t}),
\eeq
where the normalization factor is put equal to $1$.

\section{Trigonometric solutions to the KP equation}

For trigonometric solutions the tau-function has the form
\beq\label{t1}
\tau = \prod_{i=1}^{N}\Bigl (e^{2\gamma x}-e^{2\gamma x_i}\Bigr ),
\eeq
where $x_i$ depend on ${\bf t}$, so that
\beq\label{t2}
u_1=-\sum_i \frac{\gamma^2}{\sinh^2(\gamma (x-x_i))}.
\eeq
These functions have a single period $\pi i/\gamma$ in the complex plane.
It is convenient to pass to the exponentiated variables
\beq\label{t3}
w=e^{2\gamma x}, \quad w_i =e^{2\gamma x_i},
\eeq
then the tau-function becomes a polynomial with the roots $w_i$ 
and $u_1$ becomes a rational function 
with double poles at $w_i$:
\beq\label{t4}
\tau =\prod_i (w-w_i), \quad u_1= -4\gamma^2 \sum_i \frac{ww_i}{(w-w_i)^2}.
\eeq

We begin with investigation of the dynamics of poles in the time $t_2$. 
According to Krichever's method, our strategy is to solve the linear problem
(\ref{kp6}) for the $\psi$-function. 
Equation
(\ref{kp11}) suggests the following ansatz for the wave function:
\beq\label{t5}
\psi = w^{\frac{z}{2\gamma}} e^{t_1z+
t_2z^2} \left (1+ \sum_i \frac{2\gamma c_i}{w-w_i}\right ),
\eeq
where we have put the normalization factor equal to $1$ and have put
$t_k=0$ for $k\geq 3$. The coefficients $c_i$ 
depend on the times (and on $z$). 
We should substitute expressions (\ref{t4}), (\ref{t5}) into the 
linear problem
$$
-\p_{t_2}\psi +4\gamma^2 w\p_w w\p_w \psi +2u_1 \psi =0.
$$
The substitution gives:
$$
-\sum_i \frac{\dot c_i}{w-w_i}-\sum_i \frac{c_i \dot w_i}{(w-w_i)^2}+
8\gamma^2 \sum_i \frac{c_iw^2}{(w-w_i)^3}
-4\gamma z\sum_i \frac{c_iw}{(w-w_i)^2}
$$
$$
-4\gamma^2 \sum_i \frac{c_iw}{(w-w_i)^2}
-8\gamma^2 \left (\sum_i \frac{ww_i}{(w-w_i)^2}\right )
\left (\frac{1}{2\gamma} +\sum_k \frac{c_k}{w-w_k}\right )=0,
$$
where dot means the $t_2$-derivative. 
We should cancel the poles at $w=w_i$. Poles of the third order cancel automatically.
The cancellation of the second and first order poles yields the conditions
\beq\label{t6}
\left \{ \begin{array}{l}
\displaystyle{\frac{1}{2}\, \dot x_i c_i +(z-\gamma )c_i +2\gamma \sum_{k\neq i}
\frac{w_ic_k}{w_i-w_k}=-w_i}
\\ \\
\displaystyle{\dot c_i =2\gamma \dot x_i c_i -8\gamma^2 c_i \sum_{k\neq i}
\frac{w_iw_k}{(w_i-w_k)^2}+8\gamma^2 \sum_{k\neq i}\frac{w_i^2c_k}{(w_i-w_k)^2}}
\end{array}
\right.
\eeq
for $i=1, \ldots , N$ which are linear equations for the 
$c_i$'s. In a similar way, the conjugated linear problem (\ref{kp6})
with the ansatz
\beq\label{t7}
\psi^{\dag} = w^{-\frac{z}{2\gamma}} e^{-t_1z -t_2z^2} 
\left (1+ \sum_i \frac{2\gamma c^*_i}{w-w_i}\right )
\eeq
for the adjoint wave function leads to the conditions
\beq\label{t8}
\left \{ \begin{array}{l}
\displaystyle{\frac{1}{2}\, \dot x_i c^*_i +(z+\gamma )c^*_i +2\gamma \sum_{k\neq i}
\frac{w_ic^*_k}{w_k-w_i}=w_i}
\\ \\
\displaystyle{-\dot c^*_i =-2\gamma \dot x_i c^*_i -8\gamma^2 c^*_i \sum_{k\neq i}
\frac{w_iw_k}{(w_i-w_k)^2}+8\gamma^2 \sum_{k\neq i}\frac{w_i^2c^*_k}{(w_i-w_k)^2}.}
\end{array}
\right.
\eeq
It is convenient to pass to $\tilde c_i =c_i w_i^{-1/2}$, 
$\tilde c^*_i =c^*_i w_i^{-1/2}$, then the above conditions 
can be written in the matrix form
\beq\label{t9}
\Bigl ((z-\gamma )I-L\Bigr )\tilde {\bf c}=-W^{1/2}{\bf e}, \quad
\p_{t_2}\tilde {\bf c}=M\tilde {\bf c},
\eeq
\beq\label{t10}
\tilde {\bf c}^* \Bigl ((z+\gamma )I-L\Bigr )={\bf e}^TW^{1/2}, \quad
\p_{t_2}\tilde {\bf c}^*=-\tilde {\bf c}^*\tilde M,
\eeq
where $\tilde {\bf c}=(\tilde c_1, \ldots , \tilde c_N)^T$ is a column vector,
$\tilde {\bf c}^*=(\tilde c^*_1, \ldots , \tilde c^*_N)$ is a row vector,
${\bf e}=(1, 1, \ldots , 1)^T$ and the matrices $W$, $L$, $M$, $\tilde M$ are
\beq\label{t11}
W=\mbox{diag}\, (w_1, w_2, \ldots , w_N),
\eeq
\beq\label{t12}
L_{ij}=-\frac{1}{2}\, \delta_{ij}\dot x_i -2\gamma (1-\delta_{ij})\,
\frac{w_i^{1/2}w_j^{1/2}}{w_i-w_j},
\eeq
\beq\label{t13}
M_{ij}=\gamma \dot x_i \delta_{ij} -8\gamma^2 \delta_{ij}
\sum_{k\neq i} \frac{w_iw_k}{(w_i-w_k)^2}+8\gamma^2 (1-\delta_{ij})
\frac{w_i^{3/2}w_j^{1/2}}{(w_i-w_j)^2},
\eeq
\beq\label{t14}
\tilde M_{ij}=-2\gamma \dot x_i \delta_{ij}+ M_{ji} .
\eeq

It is straightforward to check the following basic commutation relation:
\beq\label{t15}
[L,W]=2\gamma \left (W^{1/2}EW^{1/2}-W\right ),
\eeq
where $E={\bf e}\otimes {\bf e}^T$ is the rank $1$ 
matrix with matrix elements $E_{ij}=1$.

The linear system (\ref{t9}) is overdetermined. A simple calculation shows that 
the compatibility condition for this system is
\beq\label{t16}
\dot L+[L,M]=0.
\eeq
We write $L=-\frac{1}{2}\dot X -A$, $M=\gamma \dot X  -2D +2B$, where
the matrices $A$, $B$, $D$, $X$ are given by
$$
A_{ik}=2\gamma (1-\delta_{ik})\frac{w_i^{1/2}w_k^{1/2}}{w_i-w_k},
\quad
B_{ik}=4\gamma^2 (1-\delta_{ik})\frac{w_i^{3/2}w_k^{1/2}}{(w_i-w_k)^2},
$$
$$
D_{ik}=4\gamma^2 \delta_{ik}\sum_{l\neq i} \frac{w_iw_l}{(w_i-w_l)^2}, \quad
X_{ik}=\delta_{ik}x_i.
$$
We have:
$$
\dot L+[L,M]=-\frac{1}{2}\ddot X -\dot A -[\dot X, B]+\gamma [\dot X, A]+
2[A,D]-2[A,B].
$$
It can be easily checked that $\dot A+[\dot X, B-\gamma A]=0$ and
$\Bigl ([A, B]-[A,D]\Bigr )_{ik}=0$ at $i\neq k$. Therefore, the compatibility
condition reduces to $\ddot X_{ii}+4[A,B]_{ii}=0$ or
\beq\label{t17}
\ddot x_i=-32 \gamma^3 \sum_{j\neq i}\frac{w_iw_j(w_i+w_j)}{(w_i-w_j)^3}=
-8\gamma^3 \sum_{j\neq i}\frac{\cosh (\gamma (x_i-x_j))}{\sinh^3 (\gamma (x_i-x_j))}.
\eeq
These are equations of motion of the trigonometric Calogero-Moser model.
Equation (\ref{t16}) is their Lax representation. It states that the evolution of the
Lax matrix $L$ is isospectral. Therefore, $H_k=\mbox{tr}\, L^k$,
$k=1, \ldots , N$, are $N$ independent integrals of motion. 
It can be proved (see \cite{Perelomov}, section 3.2) that they 
are in involution, so the system is
integrable. Introducing the momenta $p_i=\frac{1}{2}\dot x_i$, we write the 
Lax matrix (\ref{t12}) 
in the form (\ref{i2}), then the Calogero-Moser Hamiltonian $H_2$ is given by
(\ref{i3}).

\section{Dynamics of poles in higher times}

Our main tool in this section is the relation (\ref{kp7a}) which, after
substitution of the wave functions and tau-function in the form specific for
trigonometric solutions, acquires the form
$$
\frac{1}{2\pi i}\oint_{\infty}z^m \left (1+\sum_i \frac{2\gamma c_i^*}{w-w_i}\right )
\left (1+\sum_k \frac{2\gamma c_k}{w-w_k}\right )dz =4\gamma^2 \sum_i
\frac{ww_i \p_{t_m}x_i}{(w-w_i)^2}.
$$
The both sides are rational functions of $w$ with poles at $w=w_i$ vanishing at infinity.
Identifying the coefficients in front of the second order poles at $w=w_i$, we obtain
\beq\label{d1}
\p_{t_m}x_i=\frac{1}{2\pi i}\oint_{\infty}z^m\tilde c_i^* w_i^{-1}\tilde c_i \, dz.
\eeq
(Comparison of the first order poles leads to the same relation.)
From (\ref{t9}), (\ref{t10}) we conclude that
$$
\tilde {\bf c}=-(zI -(L+\gamma I))^{-1}W^{1/2} {\bf e},
\quad
\tilde {\bf c}^* ={\bf e}^T W^{1/2} (zI -(L-\gamma I))^{-1},
$$
and, therefore, (\ref{d1}) reads
\beq\label{d2}
\begin{array}{lll}
\p_{t_m}x_i &=& \displaystyle{-\res_{\infty}\sum_{k,k'}\left (
z^m w_k^{1/2}\Bigl (\frac{1}{zI-(L-\gamma I)}\Bigr )_{ki}w_i^{-1}
\Bigl (\frac{1}{zI-(L+\gamma I)}\Bigr )_{ik'}w_{k'}^{1/2}\right )}
\\ && \\
&=&\displaystyle{-\res_{\infty} \mbox{tr}\left (z^m W^{1/2}EW^{1/2}
\frac{1}{zI-(L-\gamma I)}\, W^{-1}E_i \frac{1}{zI-(L+\gamma I)}\right )},
\end{array}
\eeq
where we imply the convention $\res_{\infty}(z^{-n}) = \delta_{n1}$
and $E_i$ is the diagonal matrix with matrix elements $(E_i)_{jk}=\delta_{ij}\delta_{ik}$.
Obviously, $E_i=-\p L/\p p_i$. Using the commutation relation (\ref{t15}), we write:
\beq\label{d3}
\p_{t_m}x_i =\frac{1}{2\gamma}\, \res_{\infty}\mbox{tr}\left (z^m
(LW-WL+2\gamma W)\frac{1}{zI-(L-\gamma I)}\, W^{-1} \frac{\p L}{\p p_i}\,
\frac{1}{zI-(L+\gamma I)}\right ).
\eeq
Let us calculate
$$
\mbox{tr}\left ((LW-WL)\frac{1}{zI-(L-\gamma I)}\, W^{-1} \frac{\p L}{\p p_i}\,
\frac{1}{zI-(L+\gamma I)}\right )
$$
$$
=\mbox{tr}\! \left (W\frac{1}{zI\! -\! (L\! -\! \gamma I)}\, W^{-1} \frac{\p L}{\p p_i}\,
\frac{L}{zI\! -\! (L\! +\! \gamma I)}-\!
W\frac{L}{zI\! -\! (L\! -\! \gamma I)}\, W^{-1} \frac{\p L}{\p p_i}\,
\frac{1}{zI\! -\! (L\! +\! \gamma I)}\right )
$$
$$
=-\mbox{tr}\left (W\frac{1}{zI\! -\! (L\! -\! \gamma I)}\, W^{-1} \frac{\p L}{\p p_i}
\right )+\mbox{tr}\left (\frac{\p L}{\p p_i}\frac{1}{zI\! -\! (L\! +\! \gamma I)}\right )
$$
$$
+(z-\gamma ) \mbox{tr}\left (W\frac{1}{zI-(L-\gamma I)}\, W^{-1} \frac{\p L}{\p p_i}\,
\frac{1}{zI-(L+\gamma I)}\right )
$$
$$
-(z+\gamma )\mbox{tr}\left (W\frac{1}{zI-(L-\gamma I)}\, W^{-1} \frac{\p L}{\p p_i}\,
\frac{1}{zI-(L+\gamma I)}\right )
$$
$$
=\mbox{tr} \left (\frac{\p L}{\p p_i}\frac{1}{zI\! -\! (L\! +\! \gamma I)}-
\frac{\p L}{\p p_i}\frac{1}{zI\! -\! (L\! -\! \gamma I)}\right )
$$
$$
-2\gamma \, \mbox{tr} \left (W\frac{1}{zI-(L-\gamma I)}\, W^{-1} \frac{\p L}{\p p_i}\,
\frac{1}{zI-(L+\gamma I)}\right )
$$
(it is taken into account that the diagonal matrices $W$ and $\p L/\p p_i$ commute). 
Therefore, we get from (\ref{d3}):
\beq\label{d4}
\begin{array}{lll}
\p_{t_m}x_i &=& \displaystyle{\frac{1}{2\gamma}\res_{\infty} \mbox{tr}
\left [z^m \, \frac{\p L}{\p p_i}\left (\frac{1}{zI\! -\! (L\! +\! \gamma I)}-
\frac{1}{zI\! -\! (L\! -\! \gamma I)}\right )\right ]}
\\ && \\
&=& \displaystyle{\frac{1}{2\gamma}\, \mbox{tr} \left (\frac{\p L}{\p p_i} \, (L+\gamma I)^m -
\frac{\p L}{\p p_i} \, (L-\gamma I)^m \right )}
\\ && \\
&=& \displaystyle{\frac{1}{2(m+1)\gamma}\frac{\p}{\p p_i}\, \mbox{tr}\,
\Bigl ((L+\gamma I)^{m+1} -(L-\gamma I)^{m+1} \Bigr )=
\frac{\p {\cal H}_m}{\p p_i}},
\end{array}
\eeq
where 
\beq\label{d5}
\begin{array}{lll}
{\cal H}_m &=&\displaystyle{\frac{1}{2(m+1)\gamma}\, \mbox{tr} 
\Bigl ((L\! +\! \gamma I)^{m+1} -(L\! -\! \gamma I)^{m+1} \Bigr )}
\\&&\\
&=&\displaystyle{H_m  +
\sum_{k\geq 1} \frac{m!}{(m\! -\! 2k)! \, (2k\! +\! 1)!}\, \gamma^{2k} H_{m-2k}}.
\end{array}
\eeq
In this way we have obtained one part of the Hamiltonian equations for the higher
time flows. We see that the Hamiltonian corresponding to the $m$-th flow is a linear
combination of the Calogero-Moser Hamiltonians $H_m$, $H_{m-2}$, $H_{m-4}, \, \ldots$.
For example, ${\cal H}_3=H_3+\gamma^2 H_1$, ${\cal H}_4 =H_4+2\gamma^2 H_2 +\frac{N}{5}\gamma^4$,
${\cal H}_5=H_5+\frac{10}{3}\gamma^2 H_3 +\gamma^4 H_1$ and so on. In the case $\gamma \to 0$
(rational solutions) Shiota's result is reproduced. 

In order to obtain the remaining part of the Hamiltonian equations we differentiate
(\ref{d1}) with respect to $t_2$:
$$
\p_{t_m}\dot x_i= -2\gamma \res_{\infty}\Bigl (z^m \tilde c_i^* \dot x_i w_i^{-1}
\tilde c_i \Bigr ) +\res_{\infty} \Bigl (z^m (\tilde c_i^*  w_i^{-1}
\p_{t_2}\tilde c_i + \p_{t_2}\tilde c_i^*  w_i^{-1}
\tilde c_i )\Bigr )
$$
and use equations (\ref{t9}), (\ref{t10}). In this way we get:
$$
\p_{t_m}p_i = \frac{1}{2}\, \p_{t_m}\dot x_i =
\frac{1}{2}\sum_k \res_{\infty} \Bigl ( z^m (\tilde c_i^*  w_i^{-1}M_{ik}\tilde c_k -
\tilde c_k^* M^T_{ki}  w_i^{-1}\tilde c_i )\Bigr )
$$
$$
=-\res_{\infty}\left [ z^m \, \mbox{tr} \left ( W^{1/2}EW^{1/2}\frac{1}{zI-(L-\gamma I)}
\, G^{(i)} \frac{1}{zI-(L+\gamma I)}\right )\right ]
$$
$$
=-\frac{1}{2\gamma}\, \res_{\infty}\, \mbox{tr}\left (
z^m (LW-WL+2\gamma W)\frac{1}{zI-(L-\gamma I)}
\, G^{(i)} \frac{1}{zI-(L+\gamma I)}\right ),
$$
where the matrix $G^{(i)}$ is $G^{(i)}=\frac{1}{2}(W^{-1}E_i M -M^T W^{-1}E_i)$.
Its matrix elements are
\beq\label{d6}
G^{(i)}_{jk}=4\gamma^2 \frac{w_j^{1/2}w_k^{1/2}}{(w_j-w_k)^2}\, (\delta_{ij}-\delta_{ik}).
\eeq
A calculation similar to the one done above in this section shows that
\beq\label{d7}
\begin{array}{lll}
\p_{t_m}p_i &=&\displaystyle{-\frac{1}{2\gamma}\res_{\infty}\left [z^m \mbox{tr} 
\left (WG^{(i)} \frac{1}{zI-(L+\gamma I)}- G^{(i)}W \frac{1}{zI-(L-\gamma I)}\right )\right ]}
\\ && \\
&=&\displaystyle{-\frac{1}{4\gamma}\res_{\infty}\left [z^m \mbox{tr} \left (
(WG^{(i)}\! +\! G^{(i)}W)\left (\frac{1}{zI-(L+\gamma I)}-\frac{1}{zI-(L-\gamma I)}\right )
\right ) \right ]}
\\ && \\
&&\displaystyle{-\frac{1}{4\gamma}\res_{\infty}\left [z^m \mbox{tr} \left (
(WG^{(i)}\! -\! G^{(i)}W)\left (\frac{1}{zI-(L+\gamma I)}+\frac{1}{zI-(L-\gamma I)}\right )
\right ) \right ]}
\end{array}
\eeq
The last line here is actually equal to zero because
$$
(WG^{(i)}\! -\! G^{(i)}W)_{jk}=-2\gamma L_{jk}(\delta_{ij}-\delta_{ik})
$$
and so
$$
\mbox{tr} \Bigl ( (WG^{(i)}\! -\! G^{(i)}W)L^m\Bigr )=-2\gamma \sum_{j,l}
L_{jl}(\delta_{ij}-\delta_{il})(L^m)_{lj}
$$
$$
=-2\gamma \sum_l L_{il}(L^m)_{li}+2\gamma \sum_j (L^m)_{ij}L_{ji}=
2\gamma ((L^{m+1})_{ii}-(L^{m+1})_{ii})=0
$$
for any integer $m$. Next, it is not difficult to prove the identity
\beq\label{d8}
WG^{(i)}\! +\! G^{(i)}W =2 \frac{\p L}{\p x_i}.
\eeq
Therefore, we obtain:
\beq\label{d9}
\begin{array}{lll}
\p_{t_m}p_i &=&\displaystyle{-\frac{1}{2\gamma}\res_{\infty}\left [z^m \mbox{tr}
\left (\frac{\p L}{\p x_i}
\left (\frac{1}{zI-(L+\gamma I)}-\frac{1}{zI-(L-\gamma I)}\right )\right ) \right ]}
\\ && \\
&=&\displaystyle{-\frac{1}{2\gamma}\, \mbox{tr} \left (\frac{\p L}{\p x_i}\,
(L+\gamma I)^m -\frac{\p L}{\p x_i} \, (L-\gamma I)^m \right )}
\\ && \\
&=&\displaystyle{-\frac{\p {\cal H}_m}{\p x_i}}
\end{array}
\eeq
which is the remaining part of the Hamiltonian equations.

\section{B\"acklund transformation and equations of motion in the higher times}

Instead of dynamics of poles of the wave function one can consider equations
which connect the dynamics of poles of the wave function with dynamics of its zeros. 
This yields a transformation of the Calogero-Moser system of the B\"acklund type. 

According to (\ref{kp11}) we have
$\psi (\mu , {\bf t})=e^{x\mu +\xi ({\bf t}, \mu )}\hat \tau ({\bf t})/\tau ({\bf t})$,
where $\hat \tau ({\bf t})=\tau ({\bf t}-[\mu ^{-1}])$.
In terms of $\tau$ and $\hat \tau$ the auxiliary linear problem (\ref{kp6}) becomes
\beq\label{b0}
\p_{t_2}\log \frac{\hat \tau}{\tau}=\p_x^2 \log (\tau \hat \tau )+
\Bigl (\p_{x}\log \frac{\hat \tau}{\tau}\Bigr )^2 +2\mu \, \p_{x}\log \frac{\hat \tau}{\tau}.
\eeq
For trigonometric solutions
\beq\label{b1}
\tau =\prod_{i}\Bigl (e^{2\gamma x}-e^{2\gamma x_i}\Bigr ), \quad
\hat \tau =\prod_{i}\Bigl (e^{2\gamma x}-e^{2\gamma y_i}\Bigr )
\eeq
and in terms of the variables $w=e^{2\gamma x}$, $w_i=e^{2\gamma x_i}$, 
$v_i=e^{2\gamma y_i}$ equation (\ref{b0}) acquires the form
$$
\sum_i \frac{(\dot y_i+2\mu -2\gamma )v_i}{w-v_i} -
\sum_i \frac{(\dot x_i+2\mu +2\gamma )v_i}{w-w_i}
$$
$$
-2\gamma \sum_i \frac{w_i^2}{(w-w_i)^2}-2\gamma \sum_i \frac{v_i^2}{(w-v_i)^2}
+2\gamma \left (\sum_i \frac{v_i}{w-v_i}-\sum_i \frac{w_i}{w-w_i}\right )^2=0,
$$
where dot means the $t_2$-derivative. The second order poles cancel identically. 
Equating residues at the poles at $w=w_i$ and $w=v_i$ to zero, we get the system
of equations
$$
\left \{
\begin{array}{l}
\displaystyle{p_i=-\mu +\gamma \sum_{k\neq i}\frac{w_i+w_k}{w_i-w_k}-
\gamma\sum_{k}\frac{w_i+v_k}{w_i-v_k}}
\\ \\
\displaystyle{\tilde p_i=-\mu -\gamma \sum_{k\neq i} \frac{v_i+v_k}{v_i-v_k}+
\gamma\sum_{k} \frac{v_i+w_k}{v_i-w_k}}
\end{array}
\right.
$$
(here $p_i=\frac{1}{2}\, \dot x_i$, $\tilde p_i=\frac{1}{2}\, \dot y_i$) or
\beq\label{b2}
\left \{
\begin{array}{l}
\displaystyle{p_i=-\mu +\gamma \sum_{k\neq i}\coth (\gamma (x_i-x_k))-
\gamma \sum_{k}\coth (\gamma (x_i-y_k))}
\\ \\
\displaystyle{\tilde p_i=-\mu -\gamma \sum_{k\neq i}\coth (\gamma (y_i-y_k))+
\gamma \sum_{k}\coth (\gamma (y_i-x_k)).}
\end{array}
\right.
\eeq
Note that this is a canonical transformation 
$(p_i, x_i)\to (\tilde p_i, y_i)$ with the generating function
\beq\label{b3}
F=\sum_{i<j}\log \Bigl [ \sinh (\gamma (x_i-x_j))\sinh (\gamma (y_i-y_j))\Bigr ]
-\sum_{i,j}\log \sinh (\gamma (x_i-y_j))-\mu \sum_i (x_i-y_i)
\eeq
since we have $p_i=\p F/\p x_i$, $\tilde p_i=-\p F/\p y_i$. Equations (\ref{b2})
appeared in \cite{W82} under the name of B\"acklund transformation and in 
\cite{ABW09} under the name of self-dual equations of motion. 

Let us introduce the differential operator
\beq\label{b4}
D(\mu )=\sum_{k\geq 1}\frac{\mu^{-k}}{k}\, \p_{t_k},
\eeq
then $\hat \tau =e^{-D(\mu )}\tau$ and $y_i=e^{-D(\mu )}x_i$,
$\tilde p_i=e^{-D(\mu )}p_i$ and equations (\ref{b2}) can be written in the form
\beq\label{b5}
\left \{
\begin{array}{l}
\displaystyle{p_i=-\mu +\gamma \sum_{k\neq i}\coth (\gamma (x_i-x_k))-
\gamma \sum_{k}\coth (\gamma (x_i-e^{-D(\mu )}x_k))}
\\ \\
\displaystyle{p_i=-\mu -\gamma \sum_{k\neq i}\coth (\gamma (x_i-x_k))+
\gamma \sum_{k}\coth (\gamma (x_i-e^{D(\mu )}x_k))}
\end{array}
\right.
\eeq
(the second equation here is obtained from the second one in (\ref{b2}) by
an overall shift of times). Subtracting the two equations in (\ref{b5}), we get
\beq\label{b6}
\sum_{k}\coth (\gamma (x_i-e^{-D(\mu )}x_k))+
\sum_{k}\coth (\gamma (x_i-e^{D(\mu )}x_k))-2
\sum_{k\neq i}\coth (\gamma (x_i-x_k))=0.
\eeq
These equations comprise the generating form of all
equations of motion of the Calogero-Moser system for all higher flows with Hamiltonians
${\cal H}_k$. Namely, the equations of motion are obtained by expansion of (\ref{b5}) 
and (\ref{b6}) in powers
of $\mu$. Some details of this expansion are given in the appendix.

\section{The tau-function}

In this section we prove that the tau-function for the trigonometric
solutions to the KP hierarchy is given by the determinant formula
\beq\label{tau1}
\tau ({\bf t})=\det_{N\times N}
\left (wI-\exp \Bigl ( -\sum_{k\geq 1}t_k {\cal L}_k \Bigr )W_0\right ),
\eeq
where $W_0=W(0)$,
$$
{\cal L}_k=(L_0+\gamma I)^k -(L_0-\gamma I)^k , \quad L_0=L(0).
$$
For the proof that the eigenvalues of the matrix
$\displaystyle{\exp \Bigl ( -\sum_{k\geq 1}t_k {\cal L}_k \Bigr )W_0}$ are
$e^{2\gamma x_i}$, where $x_i$ are coordinates of the Calogero-Moser particles
(as functions of the $t_k$'s under the Hamiltonian flows with the Hamiltonians
${\cal H}_k$), see \cite{Suris}. 

Let $V$ be a diagonalizing matrix for 
$\displaystyle{\exp \Bigl ( -\sum_{k\geq 1}t_k {\cal L}_k \Bigr )W_0}$:
$$
V\exp \Bigl ( -\sum_{k\geq 1}t_k {\cal L}_k \Bigr )W_0 V^{-1}=W,
$$
where $W$ is diagonal. It is defined up to the left multiplication by a diagonal matrix.
We fix this freedom by the condition
$$
VW_0^{-1/2}{\bf e}=W^{-1/2}{\bf e}.
$$
We know that 
the matrices $W_0$, $L_0$ satisfy the commutation relation (\ref{t15}) which we write
here in the form
\beq\label{tau1a}
W_0^{-1/2}L_0W_0^{1/2}-W_0^{1/2}L_0W_0^{-1/2}=2\gamma (E-I).
\eeq
Let us prove that the matrices $W$ and $L=VL_0V^{-1}$ satisfy the same commutation relation.
We have, following \cite{Suris}:
$$
W^{-1/2}LW^{1/2}-W^{1/2}LW^{-1/2}=W^{1/2}(W^{-1}LW-L)W^{-1/2}
$$
$$
=W^{1/2}V W_0^{-1/2}\Bigl (W_0^{-1/2}L_0W_0^{1/2}-W_0^{1/2}L_0W_0^{-1/2}\Bigr )
W_0^{1/2}V^{-1}W^{-1/2}
$$
$$
=2\gamma W^{1/2}V W_0^{-1/2}({\bf e}\otimes {\bf e}^T -I)W_0^{1/2}V^{-1}W^{-1/2}
$$
$$
=2\gamma \Bigl (W^{1/2}V W_0^{-1/2}{\bf e}\otimes {\bf e}^T W_0^{1/2}V^{-1}W^{-1/2}-I\Bigr ).
$$
Since the diagonal elements of the matrix in the left hand side are equal to $0$, 
we conclude that ${\bf e}^T W_0^{1/2}V^{-1}={\bf e}^T W^{1/2}$ and obtain
$2\gamma ({\bf e}\otimes {\bf e}^T -I)$ in the right hand side.

Below we will prove that the function (\ref{tau1}) satisfies the bilinear relation
(\ref{kp9}) which we write here in the equivalent form 
\beq\label{tau2}
\begin{array}{c}
\displaystyle{
\frac{\p_x \tau ({\bf t}+[\lambda^{-1}]-[\mu^{-1}])}{\tau ({\bf t})}-
\frac{\p_x \tau ({\bf t})}{\tau ({\bf t})}\, 
\frac{\tau ({\bf t}+[\lambda^{-1}]-[\mu^{-1}])}{\tau ({\bf t})}}
\\ \\
-\, \displaystyle{
(\lambda -\mu ) \left (\frac{\tau ({\bf t}+[\lambda^{-1}]-[\mu^{-1}])}{\tau ({\bf t})}-
\frac{\tau ({\bf t}+[\lambda^{-1}])}{\tau ({\bf t})}\,
\frac{\tau ({\bf t}-[\mu^{-1}])}{\tau ({\bf t})}\right )}=0
\end{array}
\eeq
which differs from (\ref{kp9})
by a shift of the time variables.

Performing the similarity transformation with the diagonalizing matrix $V$ under the 
determinant (\ref{tau1}), we have:
$$
\tau ({\bf t})=\det (w-W),
$$
$$
\tau ({\bf t}+[\lambda^{-1}])=\det \left (wI-
\frac{(\lambda -\gamma )I-L}{(\lambda +\gamma )I-L}\, W\right )
$$
$$
=\frac{\det \Bigl (((\lambda +\gamma )I-L)
(wI-W)+2\gamma W\Bigr )}{\det \Bigl ((\lambda +\gamma )I-L\Bigr )}=
\frac{\det \Bigl ((wI-W)((\lambda +\gamma )I-L)
+2\gamma \tilde E\Bigr )}{\det \Bigl ((\lambda +\gamma )I-L\Bigr )},
$$
where $\tilde E=W^{1/2}EW^{1/2}$ and we used the commutation relation (\ref{t15}).
Using the formula $\det (I+A)=1+\mbox{tr}\, A$ valid for any rank $1$ matrix $A$, we get: 
\beq\label{tau3}
\tau ({\bf t}+[\lambda^{-1}])=
\tau ({\bf t})\left (1+2\gamma \, \mbox{tr} \Bigl [((\lambda +\gamma )I-L)^{-1}
(wI-W)^{-1}\tilde E\Bigr ]\right ).
\eeq
In a similar way, we obtain:
\beq\label{tau4}
\tau ({\bf t}-[\mu^{-1}])=\tau ({\bf t})\left (1-2\gamma \, \mbox{tr} 
\Bigl [(wI-W)^{-1}((\mu -\gamma )I-L)^{-1}
\tilde E\Bigr ]\right ).
\eeq
At last, for the tau-function $\tau ({\bf t}+[\lambda^{-1}]-[\mu^{-1}])$ we have:
$$
\tau ({\bf t}+[\lambda^{-1}]-[\mu^{-1}])=
\det \left (wI-
\frac{(\lambda -\gamma )I-L}{(\lambda +\gamma )I-L}\, 
\frac{(\mu +\gamma )I-L}{(\mu -\gamma )I-L}\,
W\right )
$$
$$
=\det \left (wI-
\frac{(\lambda -\gamma )I-L}{(\lambda +\gamma )I-L}\, W\, 
\frac{(\mu +\gamma )I-L}{(\mu -\gamma )I-L}\right )
$$
$$
=\frac{\det \Bigl (((\mu -\gamma )I-L)(wI-W)((\lambda +\gamma )I-L)-2\gamma
(\lambda -\mu )\tilde E\Bigr )}{\det \Bigl ((\lambda +\gamma )I-L\Bigr )
\det \Bigl ((\mu -\gamma )I-L\Bigr )}
$$
$$
=\tau ({\bf t})\left (1-2\gamma (\lambda -\mu )\, \mbox{tr} \Bigl [
((\lambda +\gamma )I-L)^{-1}(wI-W)^{-1}((\mu -\gamma )I-L)^{-1}\tilde E\Bigr ]\right ).
$$
Substituting everything into (\ref{tau2}), we write the left hand side of (\ref{tau2})
(divided by $\lambda -\mu$) in the form
$$
\mbox{LHS of (\ref{tau2})}\propto
2\gamma \, \mbox{tr}\left (\frac{1}{(\lambda +\gamma )I-L}\, \frac{w}{(wI-W)^2}\, 
\frac{1}{(\mu -\gamma )I-L}\, \tilde E\right )
$$
$$
+\, (\lambda -\mu ) \, \mbox{tr}\left (\frac{1}{(\lambda +\gamma )I-L}\, \frac{1}{wI-W}\, 
\frac{1}{(\mu -\gamma )I-L}\, \tilde E\right )
$$
$$
-\, \mbox{tr} \left (\frac{1}{wI-W}\, 
\frac{1}{(\mu -\gamma )I-L}\, \tilde E\right )+\, 
\mbox{tr} \left (\frac{1}{(\lambda +\gamma )I-L}\, \frac{1}{wI-W}\, \tilde E \right )
$$
$$
-\, 2\gamma \, \mbox{tr} \left (\frac{1}{wI-W}\, 
\frac{1}{(\mu -\gamma )I-L}\, \tilde E\right )
\mbox{tr} \left (\frac{1}{(\lambda +\gamma )I-L}\, \frac{1}{wI-W}\, \tilde E \right ).
$$
It is a rational function of $w$ with second and first order poles at $w=w_i$ vanishing 
at infinity. It is easy to see that poles of the second order cancel. The analysis of the
first order poles is more complicated. The residue at the pole at $w=w_i$ is equal to
$$
(\lambda -\mu +2\gamma )\sum_{j,k}\Bigl (\frac{1}{(\lambda +\gamma )I-L}\Bigr )_{ji}
\Bigl (\frac{1}{(\mu -\gamma )I-L}\Bigr )_{ik}w_k^{1/2}w_j^{1/2}
$$
$$
-\sum_j \Bigl (\frac{1}{(\mu -\gamma )I-L}\Bigr )_{ij}w_i^{1/2}w_j^{1/2}
+\sum_j \Bigl (\frac{1}{(\lambda +\gamma )I-L}\Bigr )_{ji}w_i^{1/2}w_j^{1/2}
$$
$$
\begin{array}{c}
\displaystyle{-2\gamma \sum_{j\neq i}\sum_{k}\sum_{k'}
\frac{w_i^{1/2}w_j^{1/2}}{w_i-w_j}\,
w_k^{1/2}w_{k'}^{1/2}\left [
\Bigl (\frac{1}{(\mu -\gamma )I-L}\Bigr )_{ik}
\Bigl (\frac{1}{(\lambda +\gamma )I-L}\Bigr )_{k'j} \right.}
\\ \\
\displaystyle{\left. \phantom{aaaaaaaaaaa}
+\Bigl (\frac{1}{(\mu -\gamma )I-L}\Bigr )_{jk}
\Bigl (\frac{1}{(\lambda +\gamma )I-L}\Bigr )_{k'i}\right ].}
\end{array}
$$
Since $L_{ij}$ are given by (\ref{t12}), the last triple sum is equal to
$$
\begin{array}{c}
\displaystyle{
-2\gamma \sum_{k,k'}\left [\Bigl (\frac{1}{(\mu -\gamma )I-L}\Bigr )_{ik}
\Bigl (\frac{L}{(\lambda +\gamma )I-L}\Bigr )_{k'i} \right. }
\\ \\
\displaystyle{\left. \phantom{aaaaaaaaaaaaaaaaaaaaaa}
-\Bigl (\frac{L}{(\mu -\gamma )I-L}\Bigr )_{ik}
\Bigl (\frac{1}{(\lambda +\gamma )I-L}\Bigr )_{k'i}\right ]w_k^{1/2}w_{k'}^{1/2}}
\end{array}
$$
$$
=\sum_k w_i^{1/2}w_k^{1/2}\Bigl (\frac{1}{(\mu -\gamma )I-L}\Bigr )_{ik}-
\sum_k w_i^{1/2}w_k^{1/2}\Bigl (\frac{1}{(\lambda +\gamma )I-L}\Bigr )_{ki}
$$
$$
-(\lambda -\mu +2\gamma )\sum_{k,k'}
\Bigl (\frac{1}{(\mu -\gamma )I-L}\Bigr )_{ik}
\Bigl (\frac{1}{(\lambda +\gamma )I-L}\Bigr )_{k'i}w_k^{1/2}w_{k'}^{1/2}
$$
and thus the residue is equal to $0$. We have proved that the left hand side 
of (\ref{tau2}) vanishes and, therefore, the function (\ref{tau1}) is indeed the
tau-function of the KP hierarchy.

In fact this also follows from the result of Kasman and Gekhtman \cite{KG01}
(see also \cite{Haine07}): for any square
matrices $X$, $Y$, $Z$ such that the matrix $XZ-YX$ has rank $1$ the function
\beq\label{tau5}
\tau = \det \left (X\exp \Bigl (\sum_{k\geq 1}t_k Z^k\Bigr )+\exp 
\Bigl (\sum_{k\geq 1}t_k Y^k\Bigr )\right )
\eeq
is a tau-function of the KP hierarchy. In our case $X=-W_0$, $Z=L_0-\gamma I$,
$Y=L_0+\gamma I$ and the condition that
$XZ-YX$ has rank $1$ is equivalent to the 
commutation relation (\ref{tau1a}). 
Nevertheless, we found it instructive to give an independent
direct proof. 

\section{Conclusion}

In this paper we have shown, using basically the method developed 
by Shiota for rational solutions, 
that the hierarchy of the KP equations for solutions with
trigonometric dependence on $t_1$ generates the hierarchy of the 
trigonometric Calogero-Moser dynamical equations for poles of the solutions. The KP 
hierarchical flow $t_k$ gives rise to the Hamiltonian flow with the Hamiltonian ${\cal H}_k$
which is an explicitly known linear combination of the first $k$ Hamiltonians
$H_m=\mbox{tr}\, L^m$ of the Calogero-Moser system. Therefore, there is an important 
difference with the rational case considered by Shiota, where the Hamiltonian for the 
$k$-th flow is $H_k$ itself. 

A natural unsolved problem is to extend these results to
elliptic (double periodic in the complex plane) solutions 
to the KP hierarchy whose poles are known to move
as Calogero-Moser particles with elliptic interaction potential. The $t_2$ and $t_3$ flows
are studied in \cite{Krichever80,Z19}.

\section{Appendix: expansion of equations (\ref{b5})}

Let us give some details of the expansion of equations (\ref{b5})
in powers of $\mu$. For brevity, denote
$$
c(x)=\gamma \coth (\gamma x), \quad c(x)=\frac{1}{x}+\frac{\gamma^2}{3}\, x +O(x^3)
\quad \mbox{as $x\to 0$}.
$$
We also expand
$$
e^{D(\mu )}-1=\sum_{k\geq 1}h_k(\tilde \p )\mu^{-k},
$$
where $h_k({\bf t})$ are Schur polynomials and $\tilde \p =\{\p_{t_1}, 
\frac{1}{2}\, \p_{t_2}, \frac{1}{3}\, \p_{t_3}, \ldots \}$. The first few operators
$h_k(\tilde \p )$ are:
$$
\begin{array}{l}
h_1(\tilde \p )=\p_{t_1},
\\ \\
h_2(\tilde \p )=\frac{1}{2}\, (\p_{t_2}+\p_{t_1}^2),
\\ \\
h_3(\tilde \p )=\frac{1}{6}\, (2\p_{t_3}+3\p_{t_2}\p_{t_1}+\p_{t_1}^3),
\\ \\
h_4(\tilde \p )=\frac{1}{24}\, (6\p_{t_4}+8\p_{t_3}\p_{t_1}+3\p_{t_2}^2
+6\p_{t_2}\p_{t_1}^2 +\p_{t_1}^4).
\end{array}
$$
Their action to the $x_i$'s is as follows: $h_1(\tilde \p )x_i=-1$
(because the solution essentially depends only on $x+t_1$), 
$h_2(\tilde \p )x_i=\frac{1}{2}\, \dot x_i$, 
$h_3(\tilde \p )x_i=\frac{1}{3}\, \p_{t_3}x_i$,
$h_4(\tilde \p )x_i=\frac{1}{4}\, \p_{t_4}x_i +\frac{1}{8}\, \ddot x_i$.
Expanding the second equation in (\ref{b5}), we get:
$$
p_i=-c\Bigl ((e^{D(\mu )}-1)x_i\Bigr )-\mu +\sum_{j\neq i}
\Bigl (c\Bigl (x_{ij}-(e^{D(\mu )}-1)x_j\Bigr )-c(x_{ij})\Bigr )
$$
$$
\begin{array}{l}
\displaystyle{=c\Bigl (\mu^{-1}-\sum_{k\geq 2}h_k(\tilde \p )x_i \mu^{-k}\Bigr )-\mu }
\\ \\
\displaystyle{\phantom{aaaaaaaaaa}-\sum_{j\neq i}\left [ c'(x_{ij})
\Bigl (\sum_{k\geq 1}h_k(\tilde \p )x_j\mu^{-k}\Bigr )-
\frac{1}{2}\, c''(x_{ij})\Bigl (\sum_{k\geq 1}h_k(\tilde \p )x_j\mu^{-k}\Bigr )^2+\ldots 
\right ]}
\end{array}
$$
$$
=h_2(\tilde \p )x_i+\left (h_3(\tilde \p )x_i +(h_2(\tilde \p )x_i)^2+
\sum_{j\neq i}c'(x_{ij})+\frac{\gamma^2}{3}\right )\mu^{-1}
$$
$$
\begin{array}{l}
\displaystyle{+\left (h_4(\tilde \p )x_i +2h_2(\tilde \p )x_i \, h_3(\tilde \p )x_i +
(h_2(\tilde \p )x_i)^3 \phantom{\sum_{j\neq i}}\right.}
\\ \\
\displaystyle{\left. \phantom{\int}
-\sum_{j\neq i}\Bigl ((h_2(\tilde \p )x_j c'(x_{ij})-
\frac{1}{2}\, c''(x_{ij})\Bigr )-\frac{\gamma^2}{6}\, 
h_2(\tilde \p )x_i\right )\mu^{-2} +O(\mu^{-3})}
\end{array}
$$
(here $x_{ij}=x_i-x_j$). The similar expansion of the first equation in (\ref{b5}) is
$$
p_i=-h_2(-\tilde \p )x_i+\left (-h_3(-\tilde \p )x_i +(h_2(-\tilde \p )x_i)^2+
\sum_{j\neq i}c'(x_{ij})+\frac{\gamma^2}{3}\right )\mu^{-1}
$$
$$
\begin{array}{l}
\displaystyle{+\left (-h_4(-\tilde \p )x_i +2h_2(-\tilde \p )x_i \, h_3(-\tilde \p )x_i -
(h_2(-\tilde \p )x_i)^3 \phantom{\sum_{j\neq i}}\right.}
\\ \\
\displaystyle{\left. \phantom{\int}
+\sum_{j\neq i}\Bigl ((h_2(-\tilde \p )x_j c'(x_{ij})-
\frac{1}{2}\, c''(x_{ij})\Bigr )+\frac{\gamma^2}{6}\, h_2(-\tilde \p )x_i
\right )\mu^{-2} +O(\mu^{-3}).}
\end{array}
$$
Matching the coefficients in front of powers of $\mu$, we get the relation 
$p_i=\frac{1}{2}\, \dot x_i$, the equations of motion
$\displaystyle{\ddot x_i =-4\sum_{j\neq i}c''(x_{ij})}$ and the Hamiltonian equations
$$
\p_{t_3}x_i =-3p_i^2 -3\sum_{j\neq i}c'(x_{ij})-\gamma^2=\frac{\p }{\p p_i}\,
(H_3+\gamma^2 H_1),
$$
$$
\p_{t_4}x_i =4p_i^3 +4\sum_{j\neq i}(2p_i+p_j)c'(x_{ij})+4\gamma^2 p_i=
\frac{\p }{\p p_i}\, (H_4+2\gamma^2 H_2),
$$
where
$$
\begin{array}{l}
\displaystyle{H_1=-\sum_{i}p_i,}
\\ \\
\displaystyle{H_2=\sum_i p_i^2+\sum_{i\neq j}c'(x_{ij}),}
\\ \\
\displaystyle{H_3=-\sum_i p_i^3 -3\sum_{i\neq j}p_i c'(x_{ij}),}
\\ \\
\displaystyle{H_4=\sum_i p_i^4 
+\sum_{i\neq j}\Bigl (4p_i^2+2p_ip_j\Bigr )c'(x_{ij})
+2\sum_{i\neq j\neq k} c'(x_{ij})c'(x_{jk})+
\sum_{i\neq j}(c'(x_{ij}))^2}.
\end{array}
$$
This is in full agreement with the result of section 4.

\section*{Acknowledgments}

The author is grateful to
V. Pashkov for collaboration at the early
stage of this work.  
The work on this project was supported by the 
Russian Science Foundation under grant  19-11-00275.

\end{document}